\documentclass[aps,prc,twocolumn,showpacs,preprintnumbers,nofootinbib,float,superscriptaddress,longbibliography]{revtex4-2}
\usepackage{graphicx,fancybox}
\usepackage{amsmath,amssymb}
\usepackage[colorlinks=true, pdfstartview=FitV, linkcolor=red, citecolor=blue, urlcolor=magenta]{hyperref}
\usepackage{color}
\usepackage{soul}
\usepackage{array}
\usepackage{url}
\usepackage[utf8]{inputenc}
\usepackage{lineno}
\newcommand*\patchAmsMathEnvironmentForLineno[1]{%
  \expandafter\let\csname old#1\expandafter\endcsname\csname #1\endcsname
  \expandafter\let\csname oldend#1\expandafter\endcsname\csname end#1\endcsname
  \renewenvironment{#1}%
     {\linenomath\csname old#1\endcsname}%
     {\csname oldend#1\endcsname\endlinenomath}}%
\newcommand*\patchBothAmsMathEnvironmentsForLineno[1]{%
  \patchAmsMathEnvironmentForLineno{#1}%
  \patchAmsMathEnvironmentForLineno{#1*}}%
\AtBeginDocument{%
\patchBothAmsMathEnvironmentsForLineno{equation}%
\patchBothAmsMathEnvironmentsForLineno{align}%
\patchBothAmsMathEnvironmentsForLineno{flalign}%
\patchBothAmsMathEnvironmentsForLineno{alignat}%
\patchBothAmsMathEnvironmentsForLineno{gather}%
\patchBothAmsMathEnvironmentsForLineno{multline}%
}
\usepackage[normalem]{ulem}

\newcommand{\trento}{T\raisebox{-0.3ex}{R}ENTo}
\newcommand{\pythia}{\textsc{Pythia}}

\begin{document}

\title{A modular perspective to the jet suppression from  a \\ small to large radius  in very high transverse momentum jets}


\author{Manaswini Priyadarshini}
\affiliation{School of Physical Sciences, Indian Institute of Technology Mandi, Himachal Pradesh, India}

\author{Om Shahi}
\affiliation{Department of Physics, BITS Pilani K. K. Birla Goa Campus, Goa, India}

\author{Vaishnavi Sathe}
\affiliation{P.E.S's R. S. N College of Arts and Science, Farmagudi Ponda, Goa, India}

\author{Prabhakar Palni}
\email[Corresponding author: ]{prabhakar@iitmandi.ac.in}
\affiliation{School of Physical Sciences, Indian Institute of Technology Mandi, Himachal Pradesh, India}

\begin{abstract}
In this work, we expand the scope of the JETSCAPE framework to investigate the dependence of the jet nuclear modification factor, ${R_{AA}}$, on the jet radius parameter ($R$) for broader area jet cones, going all the way up to $R$ = 1.0. This study presents a comprehensive analysis of high-${p_{T}}$ inclusive jets extending up to 1 TeV to probe the quark-gluon plasma medium at much shorter distance scales. It focuses on quenching effects observed in the quark-gluon plasma formed during Pb-Pb collisions at ${\sqrt{s_{\rm NN}}}$ = 5.02 TeV, particularly for the most-central (0-10\%) collisions. Jet-medium interactions represent a pivotal domain of both theoretical and experimental QGP studies, with various models offering different assumptions to describe these phenomena. To illustrate this modular approach, this work computes the nuclear modification factor for inclusive jets via coupling of the MATTER model (which simulates the high virtuality phase of the parton evolution) with the LBT model (which simulates the low virtuality phase of the parton evolution). Additionally, the two successful energy loss models: MARTINI and AdS/CFT are employed to characterize the jet-suppression effectively within the JETSCAPE framework. The results are compared with the experimental data from the ATLAS and CMS detectors, covering jet transverse momentum (${p_{T}}$) ranging from 100 GeV to 1 TeV for ATLAS and 300 GeV to 1 TeV for CMS. The predictions made by the JETSCAPE are consistent in the high ${p_{T}}$ range as well as for extreme jet cone sizes,  showing deviation within 10-25\%. Our major focus is on calculating the double ratio (${R^{\mathrm{R}}_{\mathrm{AA}}/R^{\mathrm{R=small}}_{\mathrm{AA}}}$) as a function of jet-R and jet-${p_{T}}$, where the experimental results align well with predictions from the JETSCAPE framework, which incorporates the multi-stage evolution of the parton shower and energy loss models.

\end{abstract}

 \maketitle


\section{Introduction}
\label{Section:Intro}
The extremely hot and dense conditions created at the start of the Big Bang led to what we now understand as the soup of deconfined state of the partons, the quark-gluon plasma (QGP)~\cite{Arnold:2002ja,Arnold:2001ba,Yang:2022nei}. The QGP is of such great interest that, to study the properties of this state of matter, experimentalists have spent decades building advanced detectors capable of recreating this extremely dense, soupy state. The Relativistic Heavy-Ion Collider (RHIC)~\cite{Lokhtin:2005px,Cao:2017hhk}
and the Large Hadron Collider (LHC)~\cite{Neufeld:2011yh,Ringer:2019rfk,Chang_2020,Luo:2021hoo} conduct heavy-ion collisions where the QGP is created for very short periods. During this time, the QGP medium has a substantial impact on the propagation and modification of the parton shower. In particular, the high-${p_{T}}$ jets produced in these heavy ion collisions undergo strong yield suppression and medium modification which are together referred to as jet quenching phenomena~\cite{Gyulassy:1999zd,Qin:2015srf,Cao:2020wlm}. Jets modification has been studied in nucleus-nucleus collisions relative to proton-proton collisions to probe the properties of the QGP via constraints from model-to-data agreement ~\cite{Appel:1985dq,Baier:1996kr,Baier:1996sk,Zakharov:1996fv}. The measurement of the nuclear modification factor for jets as well as inclusive jet fragmentation functions has revealed many important characteristics of the quark-gluon plasma~\cite{ATLAS:2014dtd,ATLAS:2019pid,STAR:2020xiv,PHENIX:2003djd}. It provides strong confirmation of the interaction of partons with the deconfined plasma, the resulting medium modifications, and the eventual transition to hydrodynamic behavior in the medium~\cite{Chang_2020}.
Since there are many conclusive studies based on jet-${R_{AA}}$~\cite{ATLAS:2012tjt,ATLAS:2018gwx,ATLAS:2022vii}, our effort here is to push the limits of the current event generators and various energy loss models for a better description of jet quenching phenomena at very high transverse momenta and broader jet cones with the multi-stage evolution of the parton shower using the Jet Energy-loss Tomography with a Statistically and Computationally Advanced Program Envelope (JETSCAPE) framework (version 3.6.5) ~\cite{Putschke:2019yrg}.

Along with these measurements, we study inclusive jet spectra for p-p and Pb-Pb by varying jet resolution parameters in the anti-${k_{T}}$ algorithm which is realized in the FASTJET software package~\cite{Cacciari:2011ma}. The inclusive jet spectrum is of significant interest due to its reduced sensitivity to hadronization effects compared to observables that involve individual final-state hadrons. Here, the area of reconstructed jet cone is defined by the jet radius parameter ${R}$. Thus, by varying ${R}$, the reconstructed jet will include different proportions of energy from the medium response and the quenched jet. We also highlight the discovery of a new sensitivity to the characteristics of the QGP and the fundamental mechanisms of jet quenching in a study for jet yield suppression versus radius ${R}$~\cite{Chien_2016}. Specifically, the dependence of jet suppression on jet-${R}$, is predicted by theoretical models based on Anti-de Sitter/Conformal Field Theory (AdS/CFT) correspondence~\cite{Hulcher_2018} and perturbative QCD~\cite{Armesto_2009}.

In this paper, we begin with calculating the jet  ${R_{AA}}$ for the Pb-Pb collisions in Section~\ref{Section:Results} and compare with the experimental data from the ATLAS as well as the CMS detector for a robust test of configuration and the overall  JETSCAPE framework. The calculations include the (2+1)D MUSIC~\cite{Schenke_2010} model for hydrodynamics, which is ideal for studying various aspects of heavy-ion collisions. Most notably, this work extends to the high-${p_{T}}$ range of jets, up to 1 TeV, which enables us to explore the structure of the QGP medium at much shorter distance scales.

We further exploit the advantages that the JETSCAPE framework offers in Section~\ref{subSection:martini-ads}, that is the coupling of several different energy loss models such as MARTINI~\cite{Schenke:2009gb} and AdS/CFT~\cite{Albacete_2008} with the MATTER~\cite{Cao:2017qpx} to explore the quenching effects in a multi-stage manner. This approach provides insight into the interaction and energy loss mechanisms in the medium concerning the virtuality of the partonic jet. Comparing the experimental data with the predictions from these different models, which handle the low virtuality phase, allows us to develop a lucid understanding of the physics governing the above models.

In Section~\ref{subSection:radius-pt-dependence}, we proceed with concrete results from the above combinations of several successful models towards the jet radius dependence studies. An important measurement is done by the CMS collaboration with respect to the jet-${R_{AA}}$ in Pb-Pb collisions at  ${\sqrt{s_{\rm NN}}}$ = 5.02 TeV at the LHC, for jet cone area covering the radii from 0.2 up to 1 ~\cite{CMS:2021vui}. This measurement gives us an intricate understanding of the behaviour and interaction strength of the high-${p_{T}}$ inclusive collimated jets in proximity to the jet axis, for $R$ = 0.2 and the energy distribution around the cone for values of radii up to 1. The current JETSCAPE framework is based on calculations of perturbative QCD and evolution in a dense medium, thus allowing us to go up to a radius of order 1. We also  report the double ratio, i.e., the ratio of ${R_{AA}}$ for a given $R$
with respect to $R$ = 0.2  as a function of jet-${p_{T}}$  and jet radius $R$. This study provides a vivid picture of the energy transactions with the QGP as the jet cone size increases, along with the trends that the JETSCAPE framework predicts.

We conclude this work in Section~\ref{Section:conclusion}, with a concise account of the current JETSCAPE framework's potential to explain the jet-${R_{AA}}$ for larger area jet cones. We also shed light on the ability of the distinct combination of models to describe the jet-${R_{AA}}$.

\subsection{SIMULATION OF ENERGY LOSS USING MULTI-STAGE APPROACH IN JETSCAPE}
\label{SubSection:simulation}
The JETSCAPE framework provides an ideal environment for carrying out multi-stage energy loss. The hard scattering is generated by \pythia\ 8~\cite{Sjostrand:2019zhc} with initial state radiation (ISR) and multiparton interaction (MPI)~\cite{PhysRevD.36.2019} enabled, and final state radiation (FSR) disabled. For the event wise simulations, the \trento\ model~\cite{Moreland:2014oya} sets up the initial conditions and the viscous hydrodynamic evolution is described by the (2+1)D MUSIC~\cite{Schenke_2010} model. This is then followed by Cooper-Frye prescription~\cite{Vovchenko_2022,McNelis_2021} which converts the fluid cells to hadrons on an isothermal hypersurface~\cite{Casalderrey-Solana:2016jvj} at ${T_\mathrm{SW}}$= 151 MeV, where ${T_\mathrm{SW}}$ is the temperature of the plasma below which particlization~\cite{Huovinen_2012} occurs at a certain hypersurface. The jet energy loss induced by scattering is calculated in a succession of two stages: MATTER~\cite{Cao:2017qpx,Majumder:2013re} which takes care of the highly virtual phase (the first stage) while the low virtuality phase (the second stage), is handled concurrently by the LBT model~\cite{Liu:2021dpm,He:2015pra,Cao:2016gvr}. We have also employed the MARTINI~\cite{Schenke:2009gb,Shi:2022rja,Yazdi:2022bru} and the AdS/CFT model~\cite{Albacete_2008} in combination with the MATTER model to explore the low virtuality phase. The virtuality of the parton is defined as ${Q}^2 = p^{\mu}p_{\mu} - m^2$. The parton undergoes energy loss in two stages, when the virtuality of the parton, ${Q^{2} > Q^{2}_{\mathrm{SW}}}$, where ${Q_{\mathrm{SW}}}$ is the switching virtuality, the MATTER model handles the energy loss and the parton is transferred to the LBT model once ${Q^{2} \leq Q^{2}_{\mathrm{SW}}}$. In these calculations, the jet medium interaction includes inelastic medium-induced gluon radiation and a medium recoil. An extensive account of the comparison of the model to the existing jet-${R_{AA}}$ is already done~\cite{JETSCAPE:2022jer}, showing that the contribution of medium recoil is quite significant in the modification of jet-${R_{AA}}$ for a complete set of centrality classes ranging from the most central collisions to the peripheral collisions. This version of the JETSCAPE framework~\cite{Putschke:2019yrg} encompasses the modifications of a Hard Thermal Loop (HTL)~\cite{Hidaka_2009} for fixed coupling ($\alpha_\mathrm{s}^\mathrm{fix}$), running coupling ($\alpha_\mathrm{s}$), and with a virtuality dependent factor, ${f(Q^{2})}$, that modulates the effective value of jet transport coefficient ($\hat{q}$). The jet transport coefficient $\hat{q}$ is the ratio of the average of the momentum squared (which is exchanged between the jet parton and the medium) to the unit interaction length traversed by the jet parton in a direction perpendicular to the momentum of the jet parton. The JETSCAPE also accounts for the reduced medium-induced emission in the high virtuality phase, arising from coherence effects. The calculations based on the HTL considering weak-coupling approximation and the limits of high-temperature yield for a $\hat{q}$ is given as~\cite{He:2015pra},
\begin{align}
\hat{q}_{\mathrm{HTL}} =C_{a}\frac{42 \zeta(3)}{\pi}  \alpha^{2}_\mathrm{s} T^{3} \ln\left[ \frac{ET}{3\pi T^{2} \alpha_\mathrm{s}^\mathrm{fix}} \right] \label{eq:HTL-qhat-formula-C-2}
\end{align}
where $C_a$ is the representation specific Casimir, $E$ is the energy of the hard parton, and $T$ is the local temperature of the medium. 
The coherence effects, which reduce the interaction strength of the parton with the medium are also taken into consideration as the virtuality-dependent  modulation factor, which regulates the effective value of 
$\hat{q}$ in the high virtuality MATTER event generator.
The virtuality-dependent modulation factor is parameterized as follows~\cite{JETSCAPE:2022jer},
\begin{align}
\hat{q} \cdot f &\!\equiv\! \hat{q}^\mathrm{run}_\mathrm{HTL}f(Q^2) ,
\label{eq:q2-dep-qhat} \\
f(Q^2) & = \left\{ \begin{array}{cc} \frac{1+10\ln^{2}(Q^2_\mathrm{sw}) + 100\ln^{4}(Q^2_\mathrm{sw})}{1+10\ln^{2}(Q^2) + 100\ln^{4}(Q^2)} & Q^2 > Q_{\rm sw}^2 \\ 1 & Q^2 \le Q_{\rm sw}^2 \end{array} \right.
\label{eq:qhatSuppressionFactor}
\end{align}
here $Q^2$ represents the running virtuality of the hard parton and $\hat{q}^\mathrm{run}_\mathrm{HTL}$ denotes the formulation with running coupling.
Finally, the partons undergo colorless hadronization according to the default Lund string fragmentation from \pythia\ 8~\cite{Sjostrand:2019zhc,Armesto:2009fj}.
The contributions in final jets are from the hard jet shower part and the effect from the soft medium response, the latter is calculated via the Cooper-Frye formula~\cite{Vovchenko_2022,McNelis_2021}. We reconstruct jets for different radius selections by applying a minimum track-particle transverse momentum requirement of ${p^\mathrm{track,min}_\mathrm{T}>}$ 0.5 GeV using the anti-${k_{T}}$ algorithm from ${\mathrm{FASTJET}}$, and then compare the results with experimental data. All the parameters involved in the tuning of the constituent models follow the standard set of tunes released by the JETSCAPE Collaboration in an elaborate recent study~\cite{JETSCAPE:2022jer}.

\begin{figure} 
\centering
\includegraphics[width=0.48\textwidth,height=8cm]{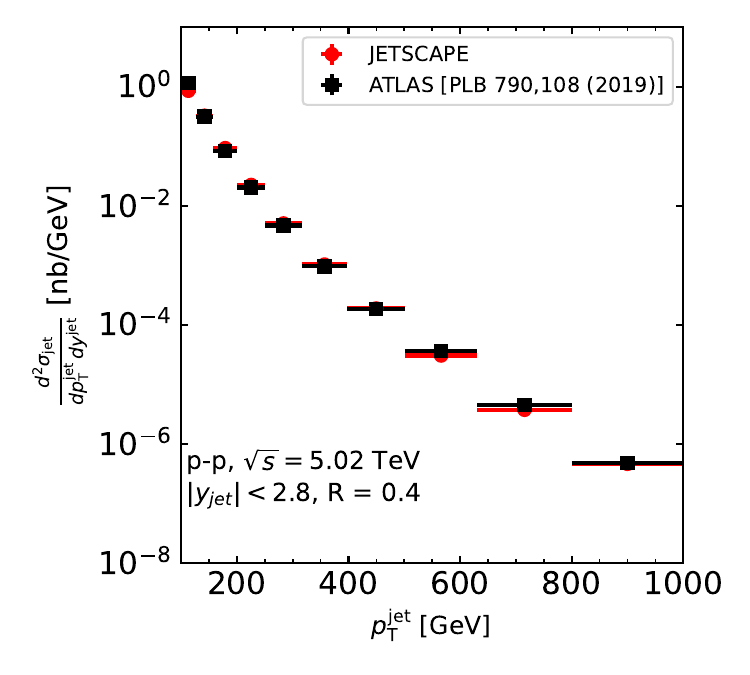}
\includegraphics[width=0.48\textwidth]{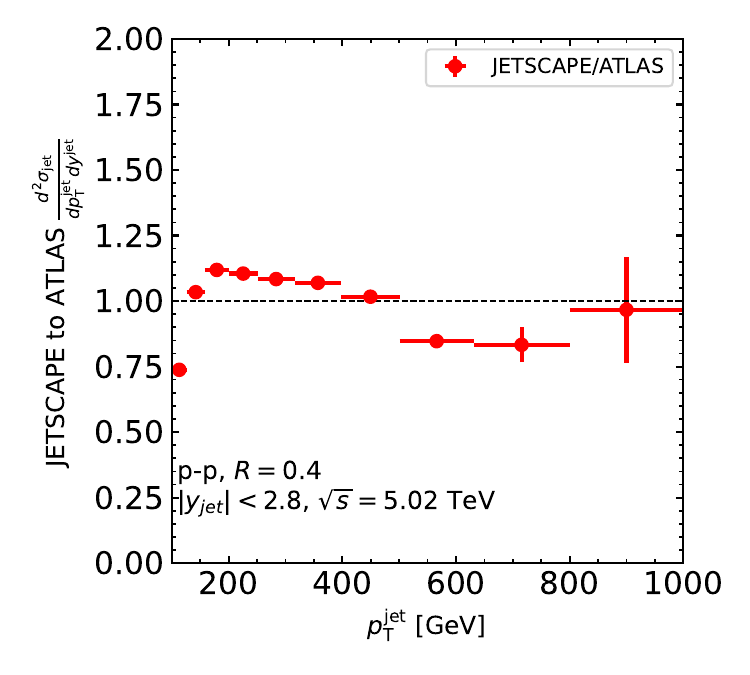}
\caption{(Color online) Differential cross-section of inclusive jets for p+p collisions at ${\sqrt{s_{\rm NN}}}$ = 5.02 TeV with the cone size $R$ = 0.4. Bottom panel shows the ratio of the JETSCAPE to the ATLAS data~\cite{ATLAS:2018gwx}.}
\label{fig:pp-spectra}
\end{figure}

\begin{figure} 
\centering
\includegraphics[width=0.48\textwidth]{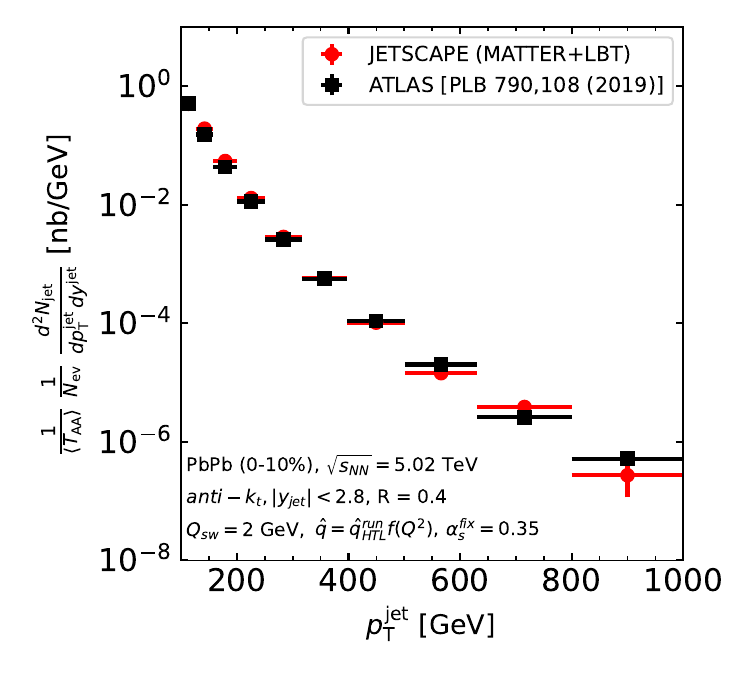}
\includegraphics[width=0.48\textwidth]{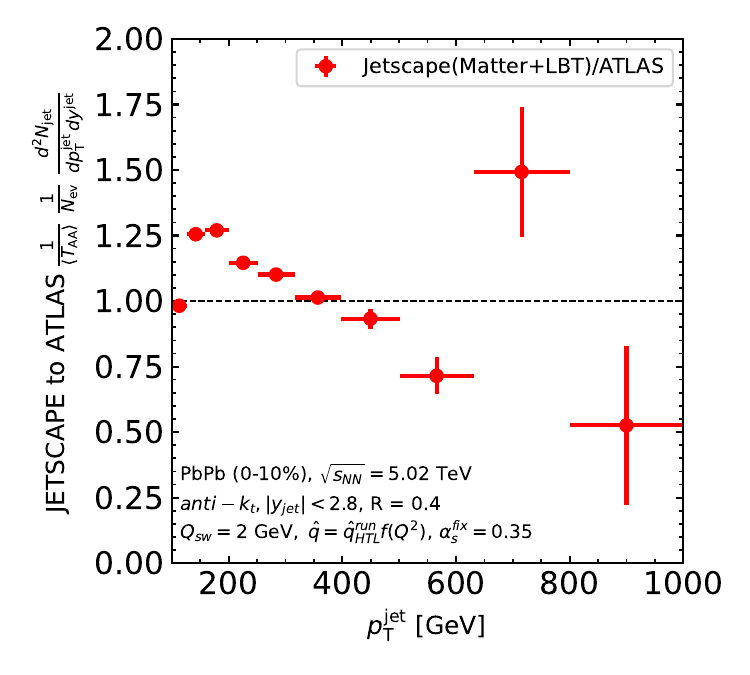}
\caption{Differential cross-section of inclusive jets in Pb+Pb collisions at ${\sqrt{s_{\rm NN}}}$ = 5.02 TeV with the cone size $R$ = 0.4. Bottom panel shows the ratio of the JETSCAPE to the ATLAS data~\cite{ATLAS:2018gwx}.}
\label{fig:pbpb-spectra}
\end{figure}

\section{Results}
\label{Section:Results}
This paper covers the collision energy of ${\sqrt{s_{\rm NN}}}$ = 5.02 TeV for the most central (0-10\%) Pb-Pb collisions and includes a comparison with selected experimental data from the ATLAS and the CMS collaborations for high-${p_{T}}$  jets. In this work, the energy loss is modeled through the coupling of the MATTER and LBT frameworks. The secondary LBT model (which handles the low virtuality phase) remains the same throughout until and unless specified. Fig.~\ref{fig:pp-spectra} (top panel) shows the ${p+p}$ collision results for inclusive jet spectra   at ${\sqrt{s}= 5.02~\rm TeV}$ for ${|y_{jet}|<2.8}$, which follow the JETSCAPE PP-19 tune~\cite{JETSCAPE:2019udz}, are then compared to the experimental data from the ATLAS~\cite{ATLAS:2018gwx}. 
In this study, we have set running coupling constant ($\alpha_\mathrm{s}$) to 0.35 for all the calculations.

The ratio of inclusive jet cross-section using the JETSCAPE to the ATLAS data is shown in the bottom panel of Fig.~\ref{fig:pp-spectra}, which shows that the results are in the acceptable range. We report the 0-10\% most central Pb-Pb jet spectra for $R$ = 0.4 in Fig.~\ref{fig:pbpb-spectra} (top panel). The ratio of the differential cross-section for inclusive jets  in Pb-Pb
collisions using JETSCAPE to the  ATLAS~\cite{ATLAS:2018gwx} data are shown in Fig.~\ref{fig:pbpb-spectra} (bottom panel). 

\vspace{2mm}

In these heavy-ion collisions, high-${p_{T}}$ jets experience energy loss due to the interactions of partons with the QGP medium. This energy loss, referred to as jet quenching, is associated with the thermodynamic properties of the QGP, leading to significant suppression and medium modification of the jets. The modification is measured using the nuclear modification factor, a key observable, which quantifies the suppression of particle yields, particularly jets, in heavy-ion collisions compared to proton-proton collisions.

\begin{align}
R_{\mathrm{AA}}
=
\frac{
\left.\frac{1}{ N_{\mathrm{evt}} }\frac{d^2 N_{\mathrm{jet}}}{dy_{jet} dp_{T}^{\mathrm{jet}}}\right|_{\mathrm{AA}}
}{
\left.\langle T_{\mathrm{AA}} \rangle \frac{d^2 \sigma_{\mathrm{jet}}}{dy_{jet} dp_{T}^{\mathrm{jet}}}\right|_{pp}
}
\end{align}

\vspace{3mm}

\begin{figure} 
\centering
\includegraphics[width=0.48\textwidth]{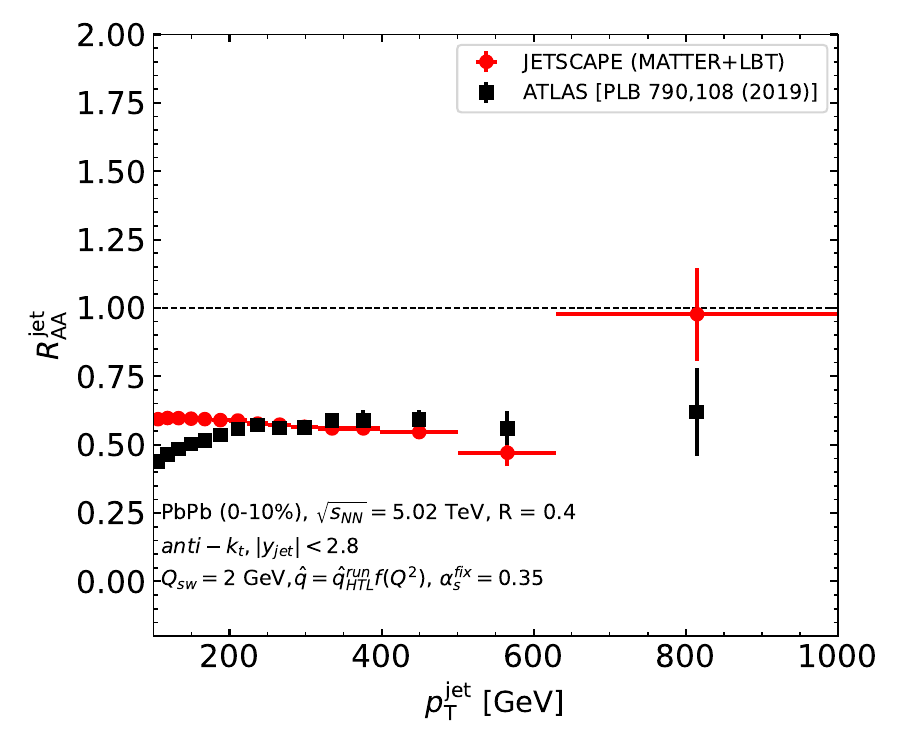}
\caption{(Color online)the jet-${R_{AA}}$ as a function of jet-${p_{T}}$ for inclusive jets in the most central (0-10\%)  Pb+Pb collisions at ${\sqrt{s_{\rm NN}}}$ = 5.02 TeV for the jet cone radius $R$ = 0.4 compared with the ATLAS data~\cite{ATLAS:2018gwx}.}
\label{fig:RAA-ATLAS}
\end{figure}

Where ${\sigma_{jet}}$ and ${N_{jet}}$ are the inclusive jet cross-section in ${p+p}$ collisions and the jet yield in Pb+Pb collisions, respectively, which are determined as a function of transverse momentum, ${p_{T}}$ and the jet rapidity, $y_\mathrm{jet}$. Moreover, ${N_{evt}}$ represents the number of Pb+Pb collisions within a specific rapidity interval and $\langle T_{\mathrm{AA}} \rangle$ denotes the ratio of the mean number of binary nucleon-nucleon collisions to the inelastic nucleon-nucleon cross-section, which can be determined using the Glauber model of the nuclear collision geometry~\cite{Miller:2007ri}.
The inclusive jet-${R_{AA}}$ is calculated as the ratio of the Pb-Pb and p-p spectra, which is presented in Fig.~\ref{fig:RAA-ATLAS}  along with a comparison to the ATLAS data~\cite{ATLAS:2018gwx}. The JETSCAPE (MATTER + LBT) results are in considerable agreement with the ATLAS measurements. However, a marginal deviation (up to ${18\%}$) is observed in the low ${p_{T}}$ region below 150 GeV, after which the results consistently align with the experimental data as ${p_{T}}$ increases.

These results validate and prove the MUSIC model's reliability, which is adequate and highly effective in describing the experimental observations even at higher jet-${p_{T}}$ in the most central collisions. In contrast to the VISHNU model~\cite{Shen:2014vra}, MUSIC's hydrodynamic evolution allows for a more accurate representation of the medium's longitudinal expansion and viscous effects. This leads to a recurring trend of consistent predictions, indicating that MUSIC outperforms other hydrodynamic models in capturing the intricate dynamics of jet interactions within the quark-gluon plasma, especially at high energy scales reaching up to 1 TeV. The MUSIC model shows low sensitivity to coherence effects that dominate at high transverse momentum, compared to VISHNU (2+1)D. This enhances the reliability of MUSIC model for analyzing the jet-suppresions phenomenon at high-${p_{T}}$.

\begin{figure*}[htbp]
\centering
\includegraphics[width=0.45\textwidth]{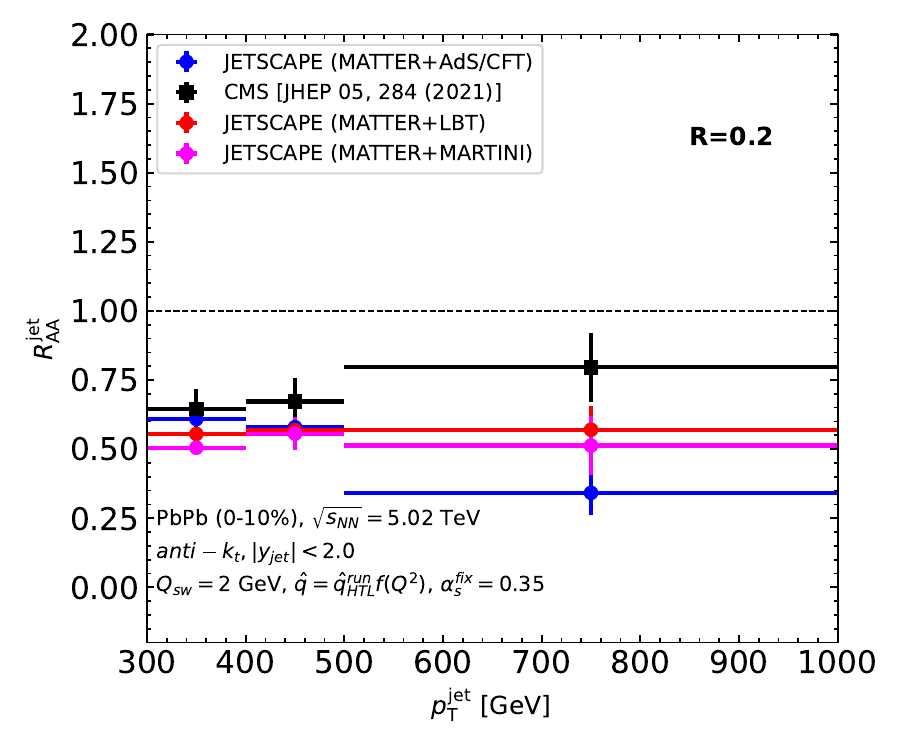}
\includegraphics[width=0.45\textwidth]{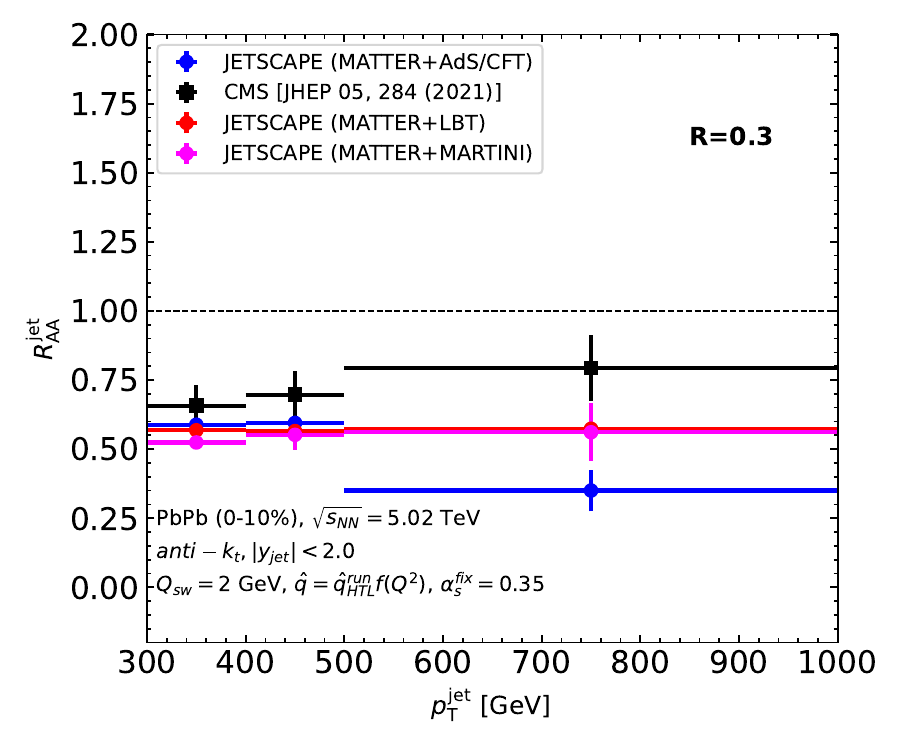}
\includegraphics[width=0.45\textwidth]{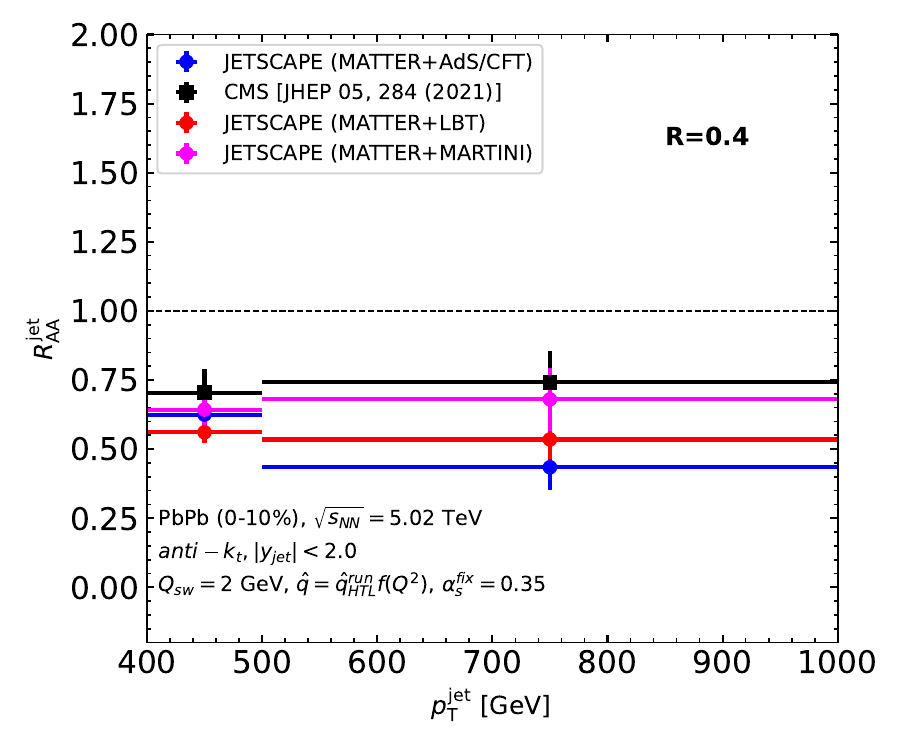}
\includegraphics[width=0.45\textwidth]{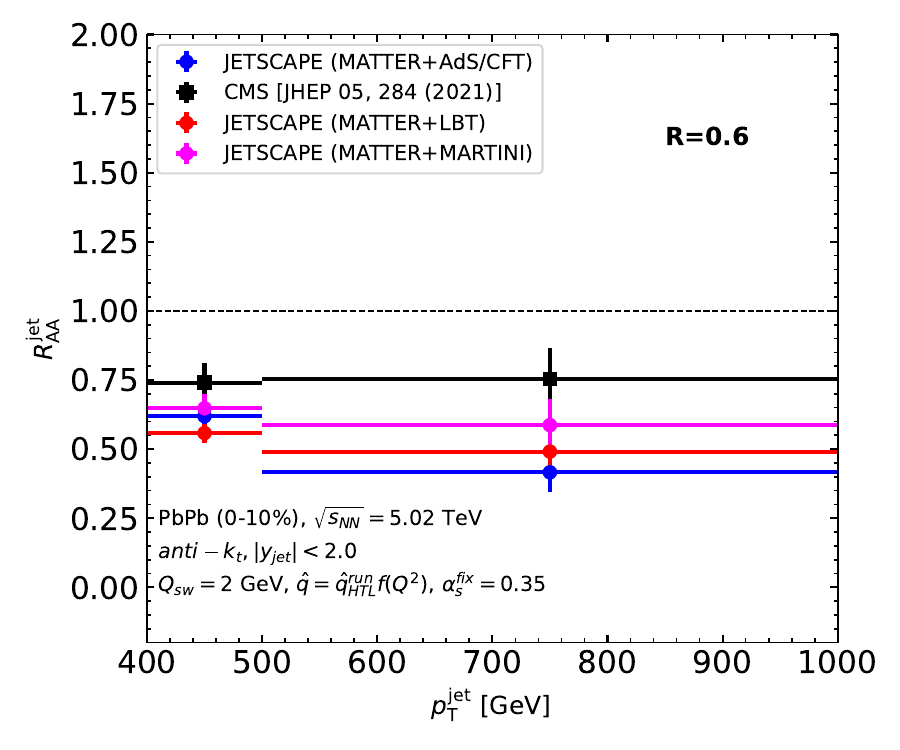}
\includegraphics[width=0.45\textwidth]{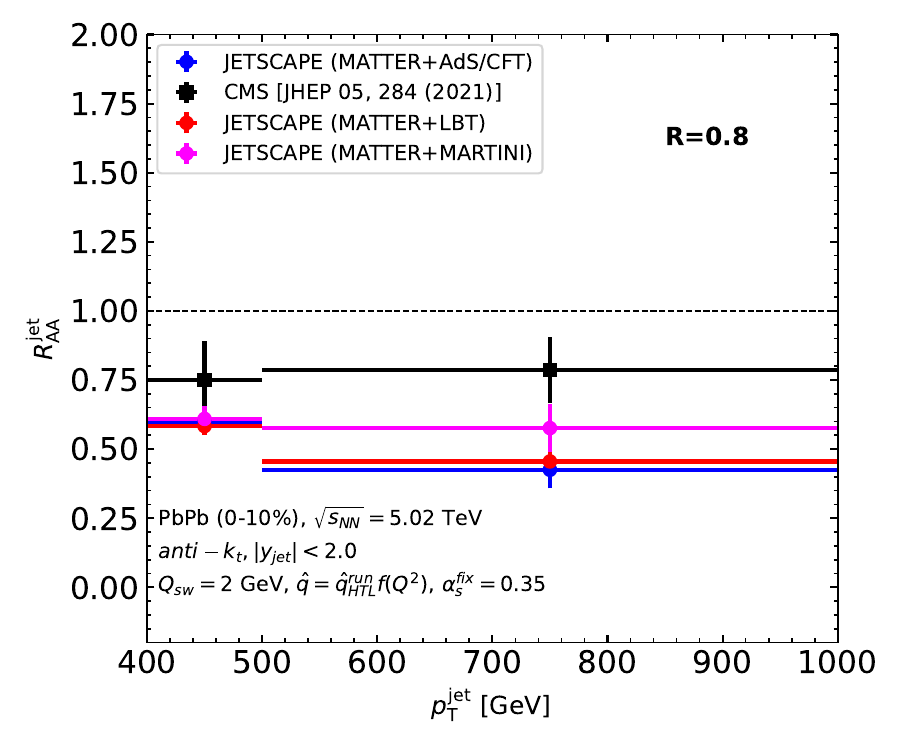}
\includegraphics[width=0.45\textwidth]{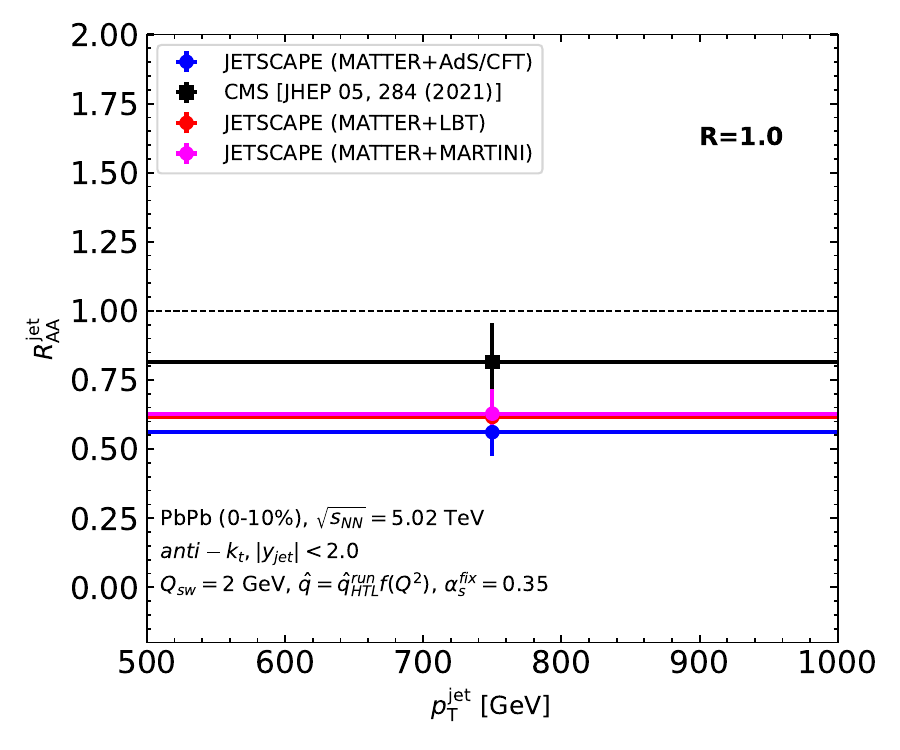}

\caption{The jet-${R_{AA}}$ as a function of jet transverse momentum ($p_{T}^{jet}$) for inclusive jets in the most central (0-10\%) Pb+Pb collisions at ${\sqrt{s_{\rm NN}}}$ = 5.02 TeV for the jet cone radii R = 0.2, 0.3, 0.4, 0.6, 0.8 and 1.0 with ${|y_{jet}|<2}$. The plot shows the comparison between the models by red, magenta, and blue markers for (MATTER + LBT), (MATTER + MARTINI), and (MATTER + AdS/CFT), respectively. The predictions are in comparison with CMS data~\cite{CMS:2021vui} shown with a black marker.}
\label{fig:RAA-CMS2}
\end{figure*}

\subsection{Exploring the MARTINI and AdS/CFT models}
\label{subSection:martini-ads}

Simulations of Pb-Pb collisions are performed using the MATTER model for both vacuum and medium showers, integrated with three distinct time-ordered parton shower models: LBT, MARTINI and AdS/CFT. The transition between these energy loss models is handled independently for each parton, with the switching parameter $Q_\mathrm{SW}$ fixed at 2 GeV. 
Both the MARTINI~\cite{Schenke:2009gb} and the AdS/CFT~\cite{Albacete_2008} models are developed to handle the low virtuality phase and simulate the energy loss once the parton is transferred from the MATTER.
We replace the successful LBT model with the MARTINI, to carry out the simulations for the same conditions~\cite{JETSCAPE:2022jer} and compare the jet-${R_{AA}}$ with the experimental data from the CMS~\cite{CMS:2021vui}. Similarly, the calculations are done by replacing the LBT with AdS/CFT model and compared with the experimental data.

This section presents the results for low virtuality parton evolution models, where the inclusive jet-${R_{AA}}$ is calculated for three setups, MATTER coupled with LBT, MATTER coupled with MARTINI, and MATTER coupled with AdS/CFT. The simulations are altogether staged in contrast with the experimental data from the CMS~\cite{CMS:2021vui}, for Pb-Pb collisions at ${\sqrt{s_{\rm NN}}}$ = 5.02 TeV and for a range of jet cone sizes up to an order of 1 and $p_{T}^{jet}$ range up to 1 TeV. 
The measurements for different models are now manifest in Fig.~\ref{fig:RAA-CMS2}, which shows the jet-${R_{AA}}$ for ${|y_{jet}|<2}$ and jet cone radius  $R$ = 0.2, $R$ = 0.3, $R$ = 0.4, $R$ = 0.6, $R$ = 0.8, and $R$ = 1.0 over a range of jet-${p_{T}}$ from 300 GeV up to 1 TeV in comparison with the experimental data from the CMS~\cite{CMS:2021vui}. The predictions made by the JETSCAPE are consistent even as we advance to the larger area jet cones. The JETSCAPE predicts marginally more suppression from a small to intermediate to large radius for all the modular combinations discussed above. The results predicted by the (MATTER + LBT) and (MATTER + MARTINI) are congruous with the experimental observations throughout the ${p_{T}}$ range and across the different jet radii. While the (MATTER + AdS/CFT) model shows more suppression at high ${p_{T}}$ region as compared to the other modular combinations.  At very high transverse momentum, the virtuality of the dominant partons is very high and the coherence effect comes into play. The coherence effect significantly reduces the interaction strength of the parton with the medium. However, AdS/CFT model is known for its strong coupling framework, which results in more significant energy loss for partons traversing the QGP medium. This leads to stronger suppression of jet-${R_{AA}}$ compared to the perturbative approaches used by LBT and MARTINI at high ${p_{T}}$ range.

\vspace{2mm}
In this section, we conclude that the JETSCAPE results depicted in the figures \ref{fig:pp-spectra}, \ref{fig:pbpb-spectra}, and \ref{fig:RAA-ATLAS} are substantial for further jet radius-dependent investigations using the current models, which is carried out in the next section.

\begin{figure*} 
\centering
\includegraphics[width=0.325\textwidth]{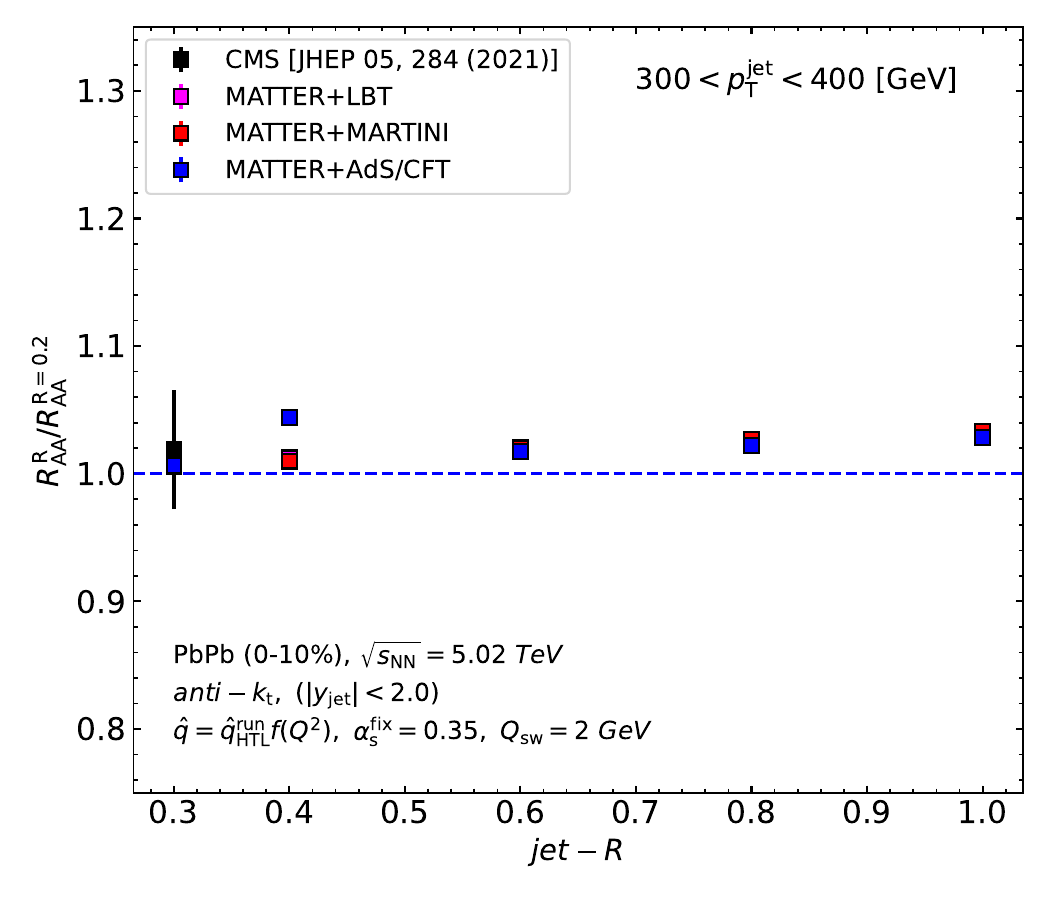}
\includegraphics[width=0.325\textwidth]{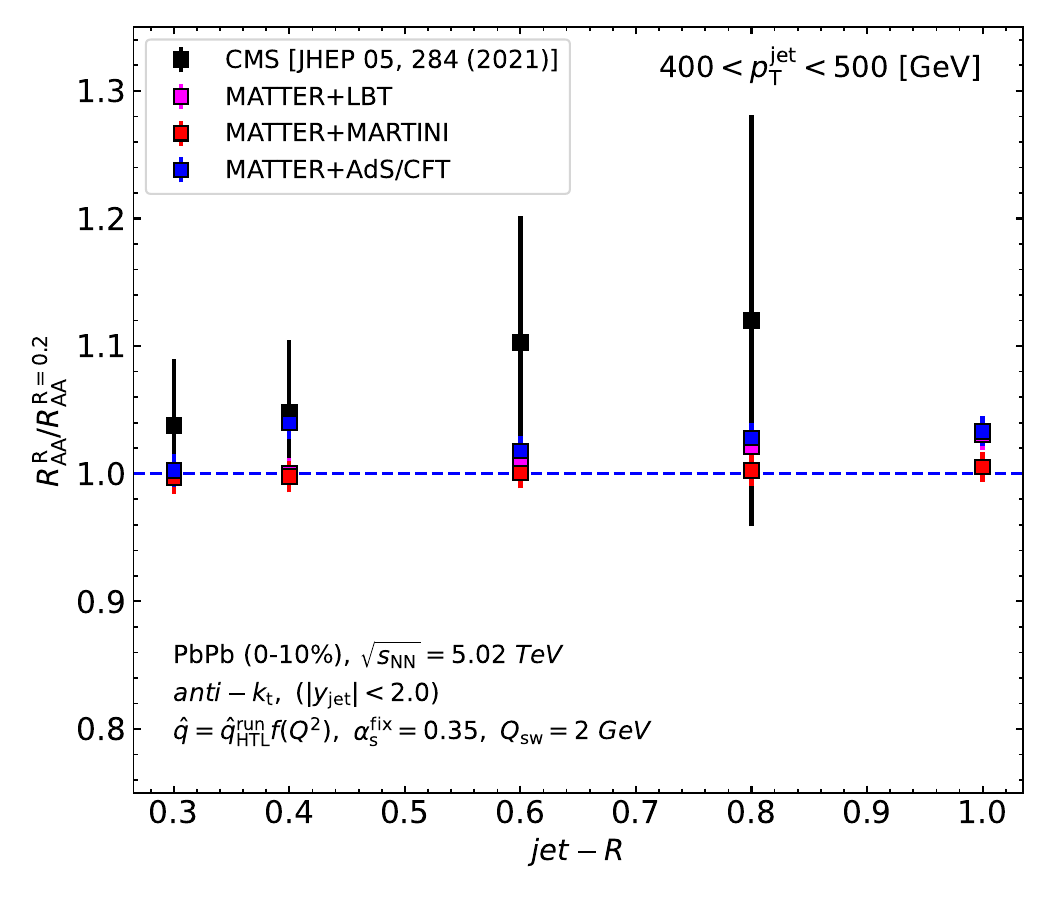}
\includegraphics[width=0.325\textwidth]{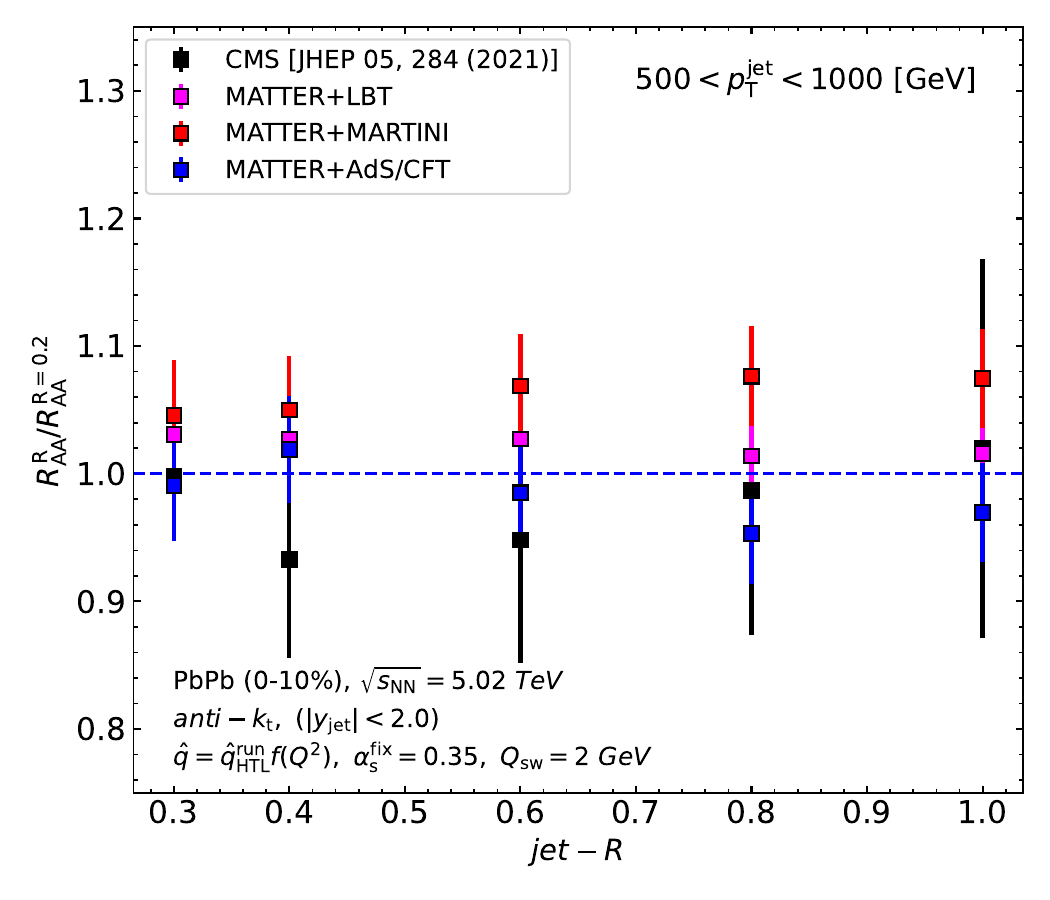}

\caption{  The double ratio (${R^{\mathrm{R}}_{\mathrm{AA}}/R^{\mathrm{R=0.2}}_{\mathrm{AA}}}$) as a function of jet radius for inclusive jets for (${\mathrm{300~GeV} \leq p_{T}^{jet} \leq \mathrm{400~GeV}}$) (left panel), (${\mathrm{400~GeV} \leq p_{T}^{jet} \leq \mathrm{500~GeV}}$) (centre panel), and  (${\mathrm{500~GeV} \leq p_{T}^{jet} \leq \mathrm{1~TeV}}$) (right panel),  in the most central(0-10\%) Pb+Pb collisions at  ${\sqrt{s_{\rm NN}}}$ = 5.02 TeV for different jet radii with ${|y_{jet}|<2}$. The plot shows the comparison between the models by magenta, red and blue markers for (MATTER+LBT), (MATTER+MARTINI), and (MATTER+AdS/CFT), respectively. The predictions are in comparison with CMS data~\cite{CMS:2021vui} shown with the black marker.}
\label{fig:double-ratioR-300-400}
\end{figure*}

\subsection{Jet radius (${R}$) and jet-${p_{\mathrm{T}}}$ dependence of the ${R_{AA}}$}
\label{subSection:radius-pt-dependence}
In this section, we emphasize the significant contribution from the hydrodynamic medium response, as already highlighted in previous studies~\cite{Chang_2020,Pablos_2020}, which is observed in the jet-${R_{\mathrm{AA}}}$ as a function of jet radius. Through radius-dependent studies, we develop a clear picture of the role played by radiation and collisions in energy loss.

A study by the CMS collaboration~\cite{CMS:2021vui} enlightens the rigorous comparisons of predictions from quenched jet event generators like HYDJET++~\cite{Lokhtin_2009}, PYQUEN~\cite{Lokhtin_2011} and theoretical models used to replicate relativistic heavy-ion collisions to the experimental data for the jet-${R_{\mathrm{AA}}}$. The article concludes with the impression that although most state-of-the-art models have progressed, there is still considerable uncertainty regarding large-area jets. The radius dependent charged-particle jet suppression analysis has also been explored in the recent ALICE study~\cite{ALICE:2023waz}. The comparison with JETSCAPE model predictions focuses on the lower jet-$p_{T}$ regime, providing valuable insight into the interplay between jet quenching and medium response for varying jet radii.

The large jet radius also indicates that the jet captures a substantial portion of the widely distributed momentum and energy deposited in the plasma. Since the JETSCAPE framework has pioneered the realization of a multi-stage approach for modular-based energy loss, the motivation here is to challenge the JETSCAPE model that has so far met our expectations in describing the variety of data observed and to investigate the limitations of the current models.

This quest is executed by calculating the jet-${R_{\mathrm{AA}}}$ double ratio (${R^{\mathrm{R}}_{\mathrm{AA}}/R^{\mathrm{R=0.2}}_{\mathrm{AA}}}$) as a function of jet radius which is in comparison with the CMS data for the most central (0-10\%) Pb-Pb collisions. The double ratio acts as a key indicator for assessing the dependence of jet energy loss on jet-$R$. When the ratio falls below unity, it indicates greater suppression of jets with larger R, while close to unity suggests no R-dependence or a balance of competing effects. Conversely, a ratio exceeding unity implies reduced suppression for jets with larger R, as demonstrated in the previous section. The results are obtained over a range of jet-${p_{T}}$ from 300 GeV up to 1 TeV and are further sub-categorized into three distinct jet-${p_{T}}$ intervals as shown in  Fig.~\ref{fig:double-ratioR-300-400}.

For ${\mathrm{300~GeV} \leq p_{T}^{jet} \leq \mathrm{400~GeV}}$ interval,  Fig.~\ref{fig:double-ratioR-300-400} (left panel) shows the predictions by the different combination of energy loss models. However, we note that there is only one measurement done by the CMS~\cite{CMS:2021vui} in this ${p_{T}}$ interval, i.e.,  for $R$ = 0.3. All three models exhibit a consistent and nearly identical trend, i.e., (MATTER + LBT), (MATTER + AdS/CFT), and (MATTER + MARTINI), each predicting a minimal enhancement in the double ratio and showing good agreement with the data.
This also indicates that the energy loss of the partons throughout the various jet cone sizes is comparable among LBT, MARTINI, and AdS/CFT.

For ${\mathrm{400~GeV} \leq p_{T}^{jet}\leq \mathrm{500~GeV}}$ interval, in Fig.~\ref{fig:double-ratioR-300-400} (center panel), we observe that the models (MATTER + LBT) and (MATTER + AdS/CFT) predict a similar trend, showing a consistent, marginal increase that aligns well with the observed data. The (MATTER + AdS/CFT) model, in particular, predicts the jet-${R_{\mathrm{AA}}}$ within ${15\%}$ of the experimental data, providing a closer match than the other models. However, the (MATTER + MARTINI) model gives double ratio values near unity in the intermediate ${p_{T}}$ range, showing a negligible R-dependence, which stands in contrast to the CMS observation.

In the extreme high $p_{T}^{jet}$ region, specifically for ${\mathrm{500~GeV} \leq p_{T}^{jet} \leq \mathrm{1~TeV}}$, shown in Fig.~\ref{fig:double-ratioR-300-400}~(right panel), the experimental data reveal that double ratio is less than unity for intermediate R. This may indicate that highly boosted jets, which possess more intricate jet substructures, undergo marginally more suppression. However, all the models moderately overestimate the jet-${R_{\mathrm{AA}}}$ double ratio in this high $p_{T}^{jet}$ region. Among them, the (MATTER + AdS/CFT) model provides the closest agreement, deviating only by ${10\%}$, and also harmonizes with the data within the uncertainties. In contrast, (MATTER + MARTINI) overpredicts the ratio, although it follows the same general pattern observed in other $p_{T}^{jet}$ regions.

\vspace{6mm}

Overall, all models predict that the double ratio (${R^{\mathrm{R}}_{\mathrm{AA}}/R^{\mathrm{R=0.2}}_{\mathrm{AA}}}$) increases with the jet-$R$, showing a systematic pattern. However, ${R_{\mathrm{AA}}}$ visibly decreases at high momentum (${p_{T}^{jet}\geq \mathrm{500~GeV} }$). In the extreme high $p_{T}^{jet}$ region (${\mathrm{500~GeV} \leq p_{T}^{jet} \leq \mathrm{1~TeV}}$), the MATTER+LBT exhibits a marginal decreasing trend at larger radii but remains above 1. The MATTER+AdS/CFT closely aligns with the CMS data, displaying a subtle decrease in the double ratio for larger jet radii. Meanwhile, the MATTER+MARTINI tends to exceed the other two models following an increasing trend in the double ratio. 

Altogether, the JETSCAPE framework convincingly describes the complete evolution of the parton shower by adopting a virtuality-based multi-stage approach for energy loss. Across the entire $p_{T}^{jet}$ regions, the radius-dependent double ratio predictions from the LBT, MARTINI and AdS/CFT models exhibit distinct jet energy loss mechanisms. MARTINI, based on the AMY formalism, combines both radiative and collisional energy loss processes. This results in a minimal medium response where energy is absorbed by the thermal background without generating diffusion wakes.
Therefore, it shows a moderate dependence on jet-$R$, with a gradual rise in double ratio as larger radii recover more radiated energy. However, the model demonstrates greater sensitivity in the higher $p_{T}$ region, where the increased contribution from gluon radiation leads to more significant changes in the double ratio. In contrast, LBT incorporates both elastic and inelastic scatterings along with a detailed treatment of thermal recoil partons and radiated gluons to capture the full jet-induced medium response. This leads to exhibit a more consistent dependence on radius compared to the other two models, with its inclusion of dynamic effects of the QGP~\cite{JETSCAPE:2018vyw}. This allows LBT to more effectively recover the energy lost outside the smaller jet cones through the medium excitation as the jet-$R$ increases. However, the AdS/CFT, which transitions to a strongly coupled energy loss mechanism characterized by continuous momentum degradation via a drag force, leads to greater suppression. As energy is predominantly dissipated into the medium with minimal recoil or recovery, this model exhibits a weak dependence on the jet radius, resulting in limited energy retention within the jet cone. Overall, these differences highlight the critical role of medium response and transport mechanisms in shaping jet quenching patterns, particularly in cone-size-dependent energy loss. They also reflect the varying assumptions regarding jet-medium interactions through all $p_{T}^{jet}$ regions, with LBT offering a more suitable framework to describe energy loss recovery. 

\begin{figure} [htbp]
\centering
\includegraphics[width=0.51\textwidth, height=7.7cm]{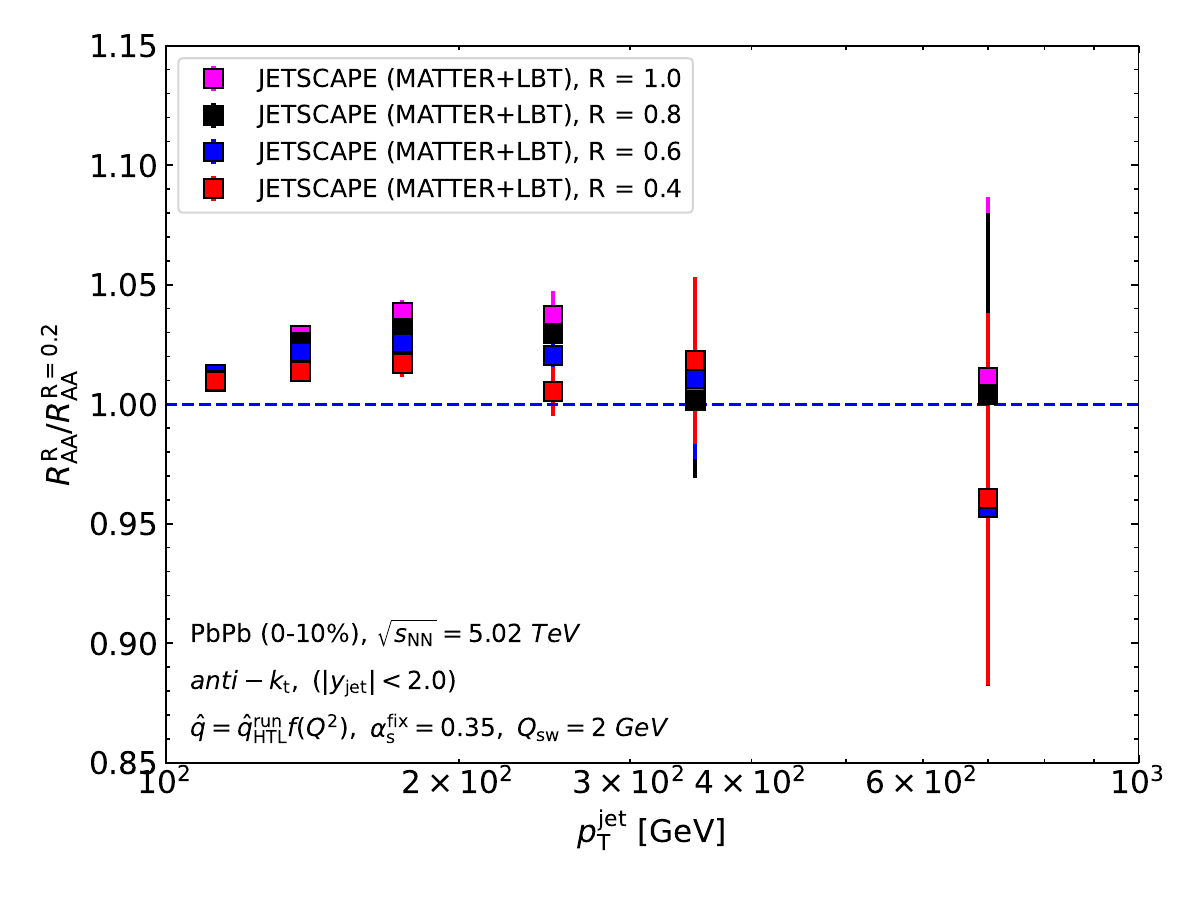}
\caption{The double ratio (${R^{\mathrm{R}}_{\mathrm{AA}}/R^{\mathrm{R=0.2}}_{\mathrm{AA}}}$) as a function of $p_{T}^{jet}$  for inclusive jets in the most central (0-10\%) Pb+Pb collisions at  ${\sqrt{s_{\rm NN}}}$ = 5.02 TeV for different jet radii.}
\label{fig:double-ratio-pT}
\end{figure}

 We extend our analysis by calculating the jet-${R_{\mathrm{AA}}}$  double ratio (${R^{\mathrm{R}}_{\mathrm{AA}}/R^{\mathrm{R=0.2}}_{\mathrm{AA}}}$) as a function of $p_{T}^{jet}$ for (MATTER + LBT) model in Fig.~\ref{fig:double-ratio-pT}. We observe a non-monotonous rise in the double ratio for $p_{T}^{jet}\leq$ 200 GeV, followed by a fall in the intermediate ${p_{T}}$ to the high ${p_{T}}$ region. The double ratio of larger radius jets is higher as compared to the smaller radius jets~\cite{ATLAS:2024jtu,ALICE:2023waz}. As the jet propagates through the medium, the final-state partons originating from jet–medium interactions are also included as jet constituents, with many of them emitted at large angles relative to the jet axis. As a result, increasing the jet radius enables the recovery of energy that would otherwise be lost outside the small jet-cone. Furthermore, the medium's response to the jet, often referred to as the wake effect~\cite{Casalderrey-Solana:2014bpa}, contributes by redirecting particles back into the jet cone. This process can lead to an increment of ${R_{\mathrm{AA}}}$ for larger R as compared with smaller R-cones. 
Therefore, as more gluons are captured within the larger jet cone, the energy lost is partially gained as the area of the jet cone increases.

\vspace{1mm}

\section{Conclusion}
\label{Section:conclusion}
In this paper, we present the jet-${R_{AA}}$ predictions from the JETSCAPE framework, which incorporates the (2+1)D MUSIC model for viscous hydrodynamic evolution and compare them with the ATLAS data for Pb+Pb collisions at  ${\sqrt{s_{\rm NN}}}$ = 5.02 TeV in jet transverse momentum interval ${\mathrm{100~GeV} \leq p_{T}^{jet} \leq   \mathrm{1~TeV}}$ for jets of radius R = 0.4. These results show that the (2+1)D MUSIC model along with (MATTER + LBT)  is successful in describing the experimental data even at higher jet-${p_{T}}$ for the most central (0-10\%) collisions. This work also elucidates the predictions made by low virtuality-based evolution models like MARTINI and AdS/CFT in a hydrodynamic medium generated by the (2+1)D MUSIC, when compared to the CMS data. We observe marginally more suppression in jet-${R_{AA}}$ in the low to intermediate  ${p_{T}}$ region.


\vspace{3mm}
We advance the current JETSCAPE calculations and compare with the data of wider jet cones ranging from $R$ = 0.2 to $R$ = 1.0 measured by the CMS detector for Pb+Pb collisions at  ${\sqrt{s_{\rm NN}}}$ = 5.02 TeV in jet transverse momentum interval ${\mathrm{300~GeV} \leq p_{T}^{jet}\leq \mathrm{1~TeV}}$. It suggests that, jets are increasingly able to recover more energy within larger radii. This can indicate that as jets propagate through the medium, more medium-induced radiation or scattered particles are incorporated within larger-radius jets.
\\

In this study, we have also shown the dependence of jet-${R_{\mathrm{AA}}}$ double ratio (${R^{\mathrm{R}}_{\mathrm{AA}}/R^{\mathrm{R=0.2}}_{\mathrm{AA}}}$) on the jet-${p_{T}}$. The analysis provides a consistent trend for (MATTER + LBT),  which shows that the double ratio is higher for jets with larger radii than for those with smaller radii. This suggests that, the larger-radius jets retain a greater fraction of the initial hard parton momentum by capturing more radiated energy within the broader cone. This leads to the reduced suppression and a higher ${R_{\mathrm{AA}}}$ compared to smaller-radius jets.
\\

This work highlights the potential of the JETSCAPE framework to describe the energy loss and medium response phenomenon for broad-area jet cones and very high transverse momentum jets. While we observe a reasonable agreement between JETSCAPE predictions and experimental data, further retuning is necessary to enhance the accuracy of jet quenching modeling in heavy-ion collisions. This improvement requires a systematic approach, incorporating next-to-leading order (NLO) jet cross-section calculations in proton-proton collisions, a key improvement as NLO corrections capture higher-order partonic effects that enhance jet-medium interactions and improve agreement with data, particularly at lower  $p_{T}$. Besides NLO corrections, precise tuning of key parameters, particularly the transport coefficient, $\hat{q}$  and strong coupling constant, $\alpha_{s}$ is essential. Along with these developments, further constraining the treatment of medium responses for large-radius jets could enhance energy redistribution and bring JETSCAPE predictions closer to experimental results.

\section{acknowledgements}
We express our gratitude to the JETSCAPE Collaboration for making the state-of-the-art framework publicly available for extensive research use. 
The authors would like to specifically thank the members of the JETSCAPE Collaboration, Yasuki Tachibana, Abhijit Majumder, Amit Kumar,  and Chun Shen for their insightful discussion and valuable feedback. The authors would  like to thank SPS local cluster facility,  and the IIT Mandi SRIC seed grant support (Ref. No. IITM/SG/2024/01-2348). The authors acknowledge National Supercomputing Mission (NSM) for providing computing resources of ‘PARAM Himalaya’ at IIT Mandi, which is implemented by C-DAC and supported by the Ministry of Electronics and Information Technology (MeitY) and Department of Science and Technology (DST), Government of India. 

\bibliography{main,misc}

\end{document}